\documentclass[aps,pra,reprint,superscriptaddress,showpacs]{revtex4-1}

\usepackage{times}
\usepackage{graphicx}
\usepackage{color}
\usepackage{amsmath}
\usepackage{amsfonts}

\newcommand{\ilac}[0]{inelastic light-assisted collisions }
\newcommand{\lac}[0]{light-assisted collisions }
\newcommand{\bs}[0]{blue Sisyphus }
\newcommand{\Rb}[0]{$^{85}$Rb}
\newcommand{\Kb}[0]{K_\textrm{B}}
\newcommand{\figref}[1]{Fig.~\ref{#1}}
\newcommand{\OmegaR}[0]{\Omega(\mathbf{r})}
\newcommand{\omegaR}[0]{\omega_1(\mathbf{r})}
\newcommand{\deltaR}[0]{\delta(\mathbf{r})}
\newcommand{\deltac}[0]{\delta_\textrm{c}}
\newcommand{\deltacmot}[0]{\delta_\textrm{c}^{33'}}
\newcommand{\ER}[0]{E(\mathbf{r})}
\newcommand{\IR}[0]{I(\mathbf{r})}
\newcommand{\zR}[0]{z_\textrm{R}}

\newcommand{\phiR}[0]{\phi(\mathbf{r})}

\newcommand{\ket}[1]{|#1\rangle}

\newcommand{\modulus}[1]{\left|#1\right|}

\newcommand{\rme}{{e}}
 \newcommand{\Fint}{$F_\textrm{int}$ }
 \DeclareSymbolFont{bbold}{U}{bbold}{m}{n}
\DeclareSymbolFontAlphabet{\mathbbold}{bbold}
\newcommand{\ind}{\mathbbold{1}}
\newcommand{\Nbar}{\bar{N}}
\begin{document}

\title{In-trap fluorescence detection of atoms in a microscopic dipole trap}
\author{A. J. Hilliard}\email[]{hilliard@phys.au.dk}
\affiliation{Dodd Walls Centre for Photonic and Quantum Technologies, University of Otago, Department of Physics, P.O. Box 56, Dunedin 9016, New Zealand.}
\affiliation{Institut for Fysik og Astronomi, Aarhus Universitet, Ny Munkegade 120,
8000 Aarhus C, Denmark.} 
\author{Y. H. Fung}
\affiliation{Dodd Walls Centre for Photonic and Quantum Technologies, University of Otago, Department of Physics, P.O. Box 56, Dunedin 9016, New Zealand.}
\author{P. Sompet}
\affiliation{Dodd Walls Centre for Photonic and Quantum Technologies, University of Otago, Department of Physics, P.O. Box 56, Dunedin 9016, New Zealand.}
\author{A. V. Carpentier}
\affiliation{Dodd Walls Centre for Photonic and Quantum Technologies, University of Otago, Department of Physics, P.O. Box 56, Dunedin 9016, New Zealand.}
\affiliation{Centro de L\'{a}seres Pulsados, Parque Cient\'{i}fico, 37185 Villamayor, Salamanca, Spain.}
\author{M. F. Andersen}
\affiliation{Dodd Walls Centre for Photonic and Quantum Technologies, University of Otago, Department of Physics, P.O. Box 56, Dunedin 9016, New Zealand.}

\begin{abstract}
We investigate fluorescence detection using a standing wave of blue-detuned light of one or more atoms held in a deep, microscopic dipole trap. The blue-detuned standing wave realizes a Sisyphus laser cooling mechanism so that an atom can scatter many photons while remaining trapped.
 When imaging more than one atom, the blue detuning limits loss due to inelastic light-assisted collisions. Using this standing wave probe beam, we demonstrate that we can count from one to  the order of 100 atoms in the microtrap with sub-poissonian precision.
\end{abstract}

\maketitle
\section{Introduction}\label{sect:Introduction}
Atomic ensembles stored in microscopic dipole traps provide an excellent system in which to study atomic collisions and light scattering processes.  These microscopic dipole traps, so-called microtraps, enable the generation of atomic samples of very high density at modest atom number and temperature.  Such a system holds the potential for generating microscopic Bose Einstein condensates~\cite{PhysRevA.88.023428}. Additionally, the use of Rydberg states within these small and dense atomic ensembles has applications in quantum information, such as the realization of collective qubits \cite{RevModPhys.82.2313} and non-linear absorption and dispersion of light through collisions between Rydberg atoms \cite{Pritchard2010,PhysRevLett.109.233602}.

One of the key challenges in this system is the determination of the number of atoms in the microtrap. In particular, determining the atom number below the level of poissonian noise is important for realizing collective quantum gates based on Rydberg ``superatoms'' \cite{RevModPhys.82.2313,PhysRevLett.112.043602}.  A standard approach is to fluorescence image the atomic sample by exposing it to red-detuned optical molasses and detecting a portion of the scattered light through high numerical aperture (NA) optics \cite{Schlosser2001,Sherson2010}.  The laser cooling provided by the optical molasses enables an atom to scatter many photons while remaining trapped. However, the optical molasses drives  light-assisted collisions that cause rapid pair-wise loss from the trap; this occurs because the red-detuned light excites a colliding pair of atoms onto an attractive molecular potential energy curve that is typically many times deeper than the optical trap \cite{Weiner2003}. This technique has been used  to produce single atom occupancy in a microtrap \cite{Schlosser2001,Aljunid2009},
but is not well-suited to the non-destructive counting of more than one trapped atom.
One can avoid this loss by releasing the atoms from the dipole trap and counting the atoms once the density is sufficiently low, either in free-space \cite{PhysRevA.82.023623}, or by recapturing them in a larger volume trap such as  a magneto-optical trap (MOT) and probing them there \cite{Hu_OptLett_19_1888_1994,Kuhr2001,PhysRevLett.95.260403,Serwane15042011}. Recently, atom counting in a MOT  has been demonstrated with single atom resolution for over 1000 atoms~\cite{PhysRevLett.111.253001}.
However, if the atomic sample is to be further manipulated within the microtrap, or if one wishes to count atoms in several adjacent microtraps,   it would be particularly useful to employ an \textit{in situ} method.

Here, we describe an in-trap fluorescence detection method that allows us to count one to more than a hundred atoms held in a microscopic dipole trap, thereby permitting subsequent probing of the trapped atomic cloud. This is an extension of earlier work~\cite{McGovern2011}, where fluorescence was induced using a standing wave of light that was blue-detuned from atomic resonance; this configuration realizes a laser cooling mechanism that works well for small detunings ($\delta\sim2\pi\times15~$MHz) and for intensities in excess of saturation, thereby yielding high photon scattering rates that make it ideal for fluorescence imaging. The sign of the detuning and the spatial variation of intensity lead to a form of Sisyphus cooling \cite{Dalibard1985,Aspect1986} that allows an atom to remain trapped during imaging and be detected with near unit efficiency. This capability permits us to eliminate the artifacts of loss during the imaging process. For samples containing more than one atom, the small  blue detuning limits the losses from inelastic light-assisted collisions, both by reducing their rate via optical shielding, and, in the case where such an inelastic collision occurs, by limiting the energy gained by the colliding atom pair to the detuning of the imaging light $\hbar\delta$ \cite{Bali1994,RevModPhys.71.1}. We show that this method can  count atoms in  samples of very high density ($n\sim 5\times10^{13}$~cm$^{-3}$) with sub-poissonian precision.

The paper is organized as follows. In the next section we  describe the experimental setup. In section~\ref{sect:LightShifts}, we summarize the methods used to calculate the light shifts of the dipole trap and probe light. Section~\ref{sect:Singleatomdetection} describes fluorescence detection of a single atom held in a microtrap, and section~\ref{sect:Multipleatomdetection} extends this to the case of multiple atoms. In section~\ref{sect:Conclusion} we present conclusions and provide an outlook for future work.

\section{Experimental setup}\label{sect:Experimentalsetup}
\begin{figure}[t!]
\includegraphics[width=8.6cm]{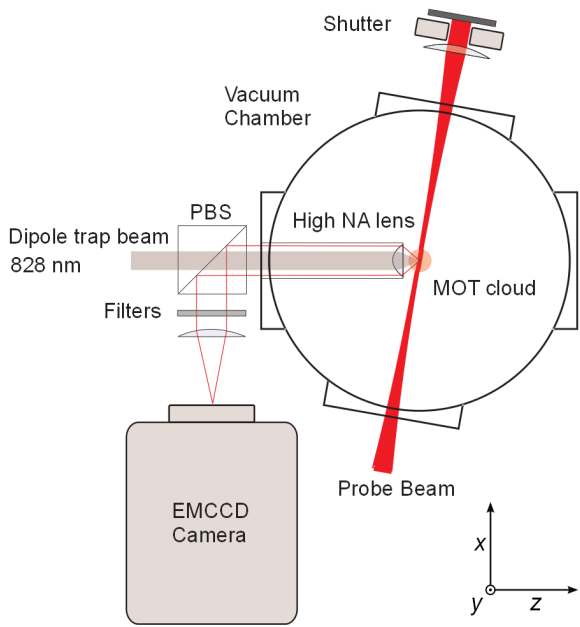}
\caption{(Color online) Schematic of the experimental setup. $^{85}$Rb atoms are loaded from a MOT into a microscopic dipole trap. The microtrap is formed by passing 828~nm light through a high NA lens mounted inside the vacuum chamber. Atoms held in the microtrap are probed by a standing wave of light on the D1 line of $^{85}$Rb.
The MOT cooling beams are not shown. A portion of the light scattered by the trapped atoms is detected by an EMCCD camera.}
\label{fig:setup}
\end{figure}
We prepare small atomic samples by loading a microscopic dipole trap from a cloud of $^{85}$Rb atoms held in a magneto-optical trap. Figure~\ref{fig:setup} is a schematic of the setup. The MOT operates on the $F=3$ to $F'=4$ transition of the D2 line of $^{85}$Rb ($5^2\mathbf{S}_{1/2}$ to $5^2\mathbf{P}_{3/2}$) at wavelength $\lambda=780$~nm. The light for the MOT is supplied as three retro-reflected  beams with gaussian intensity profiles, each with spot size ($1/\rme^2$ waist) 6.4~mm and apertured to 7~mm in diameter; throughout the paper, we state the power measured after the apertures. A compressed MOT (CMOT) phase typically lasting 150~ms overlaps the center of the MOT cloud with the dipole trap. Subsequently, the magnetic field is extinguished and  5~ms of polarization gradient cooling leads to the loading of approximately one hundred atoms in the microtrap. The number of loaded atoms  is typically varied by changing the duration of the MOT loading time ($\sim1$~s) and the CMOT phase. Three sets of Helmholtz coils serve to null the ambient magnetic field.

The microtrap is formed by focusing a linearly polarized gaussian beam with wavelength 828~nm  by a high numerical aperture lens held in the vacuum chamber to a spot size of \mbox{$w_0 = 1.8~\mu$m.} The lens  is a single-piece molded asphere with numerical aperture 0.55 and working distance 2.92~mm (Lightpath model 352230). The size and  gaussian intensity distribution of the dipole trap beam were confirmed in a replica setup outside the vacuum chamber.
These dipole beam parameters generate microscopic optical traps that are several mK deep using $\sim10$~mW of optical power; a detailed description of the light shifts induced by the dipole trap beam is presented  in section \ref{sect:LightShifts}.

To detect atoms in the microtrap, we induce fluorescence with a probe beam on the D1 line of  $^{85}$Rb. Typically, the probe beam is blue-detuned
from the $F=2$ to $F'=3$ transition on the D1 line ($5^2\mathbf{S}_{1/2}$ to $5^2\mathbf{P}_{1/2}$) for an atom at the center of the dipole trap.  We choose the D1 line for imaging so that we can block stray light from the MOT and dipole trap using standard optical filters. The detuning is given by $\delta=\omega-\omega_0$, where $\omega$ is the frequency of the laser field, and $\omega_0$ the frequency of the atomic resonance; a blue detuning corresponds to $\delta>0$. The probe beam detuning an atom experiences depends on its position in the microtrap. In this paper, we  reference the probe detuning to that experienced by an atom at the center of the dipole trap, and denote this by $\delta_\textrm{c}$. The probe beam is linearly polarized (along direction $y$ in \figref{fig:setup}) and retro-reflected to form a standing wave by a metal mirror placed outside the vacuum chamber. A mechanical shutter lies in the return beam path so that the probe beam can be toggled between traveling and standing wave configurations within the course of an experimental run. The probe beam has a gaussian intensity profile and is focused with a waist of $w_0 \cong 150~\mu$m at the position of the atoms. We use a focused beam primarily to avoid it hitting the high NA lens so as to minimize stray light in images. Due to the available ports of the vacuum chamber, the probe beam runs at an angle of  $9^\circ$ to the radial axis of the dipole trap.

Given there is no cycling transition on the D1 line, we require repump light to return the atoms to the $F=2$ ground state. For this, we primarily use the MOT cooling beams detuned by an amount $\delta_\textrm{c}^{33'}$ from the D2 $F=3$ to $F'=3$ transition for an atom at the center of the trap, averaged over the magnetic sublevels and assuming a uniform population distribution across these levels. We may also use the $F=3$ to $F'=2$ transition on the D1 line to optically pump the atoms back to the lower ground state. This beam shares the same optical path as the probe beam; i.e.,  both D1 beams propagate through the same optical fiber. Since our imaging system detects all D1 light, this repump beam increases the signal  by about 80\%~\cite{McGovern2011}.
For imaging, the D1 and D2 beams are applied with rectangular pulse envelopes.

A portion of the scattered light by the atomic sample is collected by the high NA lens and directed onto an Electron Multiplying Charge Coupled Device (EMCCD) camera.
Assuming fluorescence into a solid angle of $4\pi$, the high NA lens collects 10\% of the scattered light, and half of this is reflected by a Polarizing Beam Splitter (PBS). The remaining light is optically filtered for dipole trap light at 828~nm and D2 light at 780~nm. Room light is further suppressed using shields that enclose the vacuum chamber and surrounding optics. An image is formed on the EMCCD by a high quality lens with focal length 200~mm. The combined transmission of the PBS, filters and additional optics is 37$\%$. The EMCCD has a measured quantum efficiency of $\eta=60\%$, so that the total detection efficiency is 2.3$\%$. We typically operate the camera with multiplication gain set such that the total (mean) gain $G=\eta g=20$, where the number of analog-to-digital units (ADU) and the incident number of photons are linked by $N_\textrm{ADU}=GN_\textrm{ph}$. The gain of the EMCCD is a result of several amplification stages in series leading to an additional multiplicative noise term of approximately $\sqrt{2}$ \cite{Hynecek2003,Robbins_IEEE_50_1227_2003}.
The imaging system has magnification 44, and the camera has pixel-size 16$\times 16~\mu$m$^2$. The image of the atomic sample is integrated over an 11$\times 11$~pixel region of interest, from which we subtract the mean background level from two ($\sim$ 80 x 80 pixel) regions on either side of the atomic signal \footnote{The stochastic gain of the EMCCD leads to the fact that these `dark' regions, rather than showing a symmetric histogram of counts centered around a mean background level, instead have an asymmetric histogram characteristic of an extreme value distribution, where the tail of the distribution is longer towards higher counts. We take the  peak value of this histogram, i.e., the mode of the distribution, as the background level we subtract from the atomic signal. This leads to a small, positive bias on our images.}.

The number of atoms held in the trap can be reduced by engaging the probe beam  further blue-detuned from resonance \cite{Weiner2003}. This technique excites an atom pair onto a repulsive molecular potential curve; the energy gained by the atom pair in the process is at maximum $\hbar\delta$, in contrast to the case of excitation by a red-detuned beam. 
In previous work, we have  achieved a single atom loading efficiency in excess of 90\%~\cite{Grunzweig2010,AliciaSingle}. In this paper, we study the imaging of single atoms produced by this technique, but we also employ it to obtain a single atom fluorescence signal as a reference level when imaging multiple trapped atoms. 

\section{Light shifts}\label{sect:LightShifts}
The use of a  deep dipole trap to hold an atomic sample means that the atoms can scatter many photons and remain trapped, but it also requires us, when imaging, to take account of the significant light shifts induced by the microtrap  \cite{PhysRevA.87.063408}.
Furthermore,  the \bs mechanism used for probing relies on  the spatial variation of the light shifts induced by the standing wave probe beam,  so an accurate calculation of these shifts is critical to analyzing this mechanism  applied to in-trap fluorescence detection.
Our approach is to calculate the light shifts generated by the dipole trap and to include these as effective spatially dependent detunings in the calculation of the light shifts due to the probe beam. This approach is valid and convenient because the large detuning from any optical transition of the dipole trap light means that the dressed states within the dipole trap are very well approximated by the bare atomic states with shifted energy levels \cite{Grimm2000}.

We calculate the light shifts induced by the dipole trap in a semi-classical picture where an applied laser beam is represented by a classical electric field $\ER\cos(\omega t)$. To second order in perturbation theory, the shift of atomic level $\ket{i}$ is given by \cite{Shore1990}
\begin{equation}\label{eqn:deltaEi}
	\Delta E_i(\mathbf{r})=\frac{\modulus{\ER}^2}{4\hbar}\sum_{j\neq i}\modulus{M_{ij}}^2\left(\frac{1}{\omega-\omega_{ij}}-\frac{1}{\omega+\omega_{ij}}\right),
\end{equation}
where the sum is taken over all other electronic states $\ket{j}$, and $M_{ij}$ ($\omega_{ij}$) denotes the dipole matrix element (atomic transition frequency) between states $\ket{i}$ and $\ket{j}$.
If the frequency of the applied field is close to a given atomic transition, the rotating-wave approximation may be applied and the second term in parenthesis in Eq.~\eqref{eqn:deltaEi}  may be discarded, leaving the denominator as the detuning from the chosen transition \mbox{$\delta=\omega-\omega_{ij}$}.

To calculate the light shifts for \Rb, we use values for the dipole matrix elements and bare atomic transition frequencies \cite{Steck} to perform the sum in Eq.~\eqref{eqn:deltaEi} over the relevant dipole allowed transitions on the D1 and D2 lines~\cite{[{We neglect the contributions from other  spectral lines in \Rb, of which the strongest are  the $5^2\mathbf{S}_{1/2}$ to $6^2\mathbf{P}_{1/2}$ and $6^2\mathbf{P}_{3/2}$ lines at $\sim 420$~nm, and the $5^2\mathbf{P}_{1/2}$ and $5^2\mathbf{P}_{3/2}$ to $4^2\mathbf{D}_{3/2}$ lines at 1475~nm and 1529~nm respectively. This is a very good approximation given how close the wavelength of the dipole trap  is to those of the $5^2\mathbf{S}_{1/2}$ to $5^2\mathbf{P}_{1/2}$ and $5^2\mathbf{P}_{3/2}$ lines, and the fact that these transitions are at minimum a factor of 45 stronger than the next most intense. See }] Sansonetti}. 
The electric field strength is related to the intensity by \mbox{$\IR=\frac{1}{2}c\epsilon_0\modulus{\ER}^2$.}  The  gaussian intensity distribution of the dipole trap beam  is \mbox{$\IR=I_0\exp(-2r^2/w(z)^2)$,}
where 
$I_0=2P/(\pi w(z)^2)$, 
$P$ is the beam power, and the position dependent spot size is given by
\mbox{$w(z)=w_0\left[1+(z/\zR)^2\right]^\frac{1}{2}$}, with \mbox{$\zR=\pi w_0^2/\lambda$}  the Rayleigh range.

\begin{figure}[t]
\includegraphics[width=8.6cm]{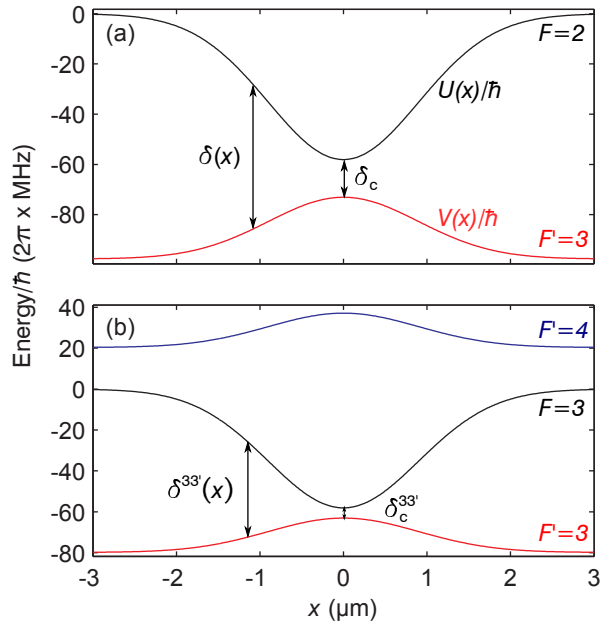}
\caption{(Color online) Light shifts for relevant states of $^{85}$Rb along the radial direction of the microtrap at a power of 20~mW. 
(a) States addressed by D1 probe beam: Ground state $F=2$ (black line) and excited state $F'=3$ (red line). The probe beam optical frequency $\omega$ has been subtracted from the excited state energy.  $\deltac$ is the detuning of the probe beam at the center of the microtrap. Away from the center, an atom experiences a position-dependent detuning $\delta(x)$ that is larger than $\deltac$.  
(b) States addressed by the D2 MOT cooling beams: Ground state $F=3$ (black line) and excited states $F'=3$ (red line) and $F'=4$ (blue line). The MOT cooling beams optical frequency $\omega_\textrm{D2}$ has been subtracted from the excited state energies.
 The magnetic sublevels in the excited states  on the D2 line   experience different light shifts; the curves show the mean across the magnetic sublevels. $\deltacmot$ denotes the detuning  of the MOT cooling beams from the $F=3 \rightarrow F'= 3$ transition on the D2 line at the center of the microtrap,  with $\delta^{33'}(x)$  the position dependent detuning.}
\label{fig:trap_lightshifts}
\end{figure}

Figure~\ref{fig:trap_lightshifts} shows   light shifts calculated using Eq.~\eqref{eqn:deltaEi} for  relevant states in $^{85}$Rb along the radial direction of the dipole trap beam. The red detuning ($\delta<0$) of the dipole trap beam shifts the ground states down and the excited states up. Figure~\ref{fig:trap_lightshifts}(a) shows the light-shifted energies of the states addressed by the D1 probe beam.
Due to the use of linearly polarized light, the magnetic sublevels of the ground states $F=2, 3$ and the excited states $F'=2, 3$ on the D1 line are shifted equally. 
 We denote the energy of the lower ground state  $U(\mathbf{r})$ and the $F'=3$ excited state on the D1 line $V(\mathbf{r})$. For 20 mW power in the dipole trap beam, \mbox{$[V(0)-U(0)]/\hbar=2\pi\times82.7$~MHz.} The microtrap leads to a position dependent detuning \mbox{$\deltaR=\delta-[V(\mathbf{r})-U(\mathbf{r})]/\hbar$} for the probe beam. 
 Figure~\ref{fig:trap_lightshifts}(b) shows the light-shifted energies of the states addressed by the MOT cooling beams on the D2 line:
 The $F=3$ ground state and the  \mbox{$F'= 3, 4$} excited states.  The light shifts for the excited states on the D2 line are $m_F$-level dependent; the curves show the mean over the magnetic sublevels, assuming a uniform population distribution across them. As for the probe beam, the microtrap leads to a position dependent detuning  for the MOT cooling light from the $F=3$ to $F'=3$ transition, $\delta^{33'}(\mathbf{r})$. However, on the D2 line, the situation is complicated by the fact that as the atom moves away from the center of the trap, the $F=3$ to $F'=4$ transition also becomes near-resonant.

To calculate the energy shifts of the atom due to the probe beam we use the dressed atom approach \cite{Dalibard1985}. For fluorescence imaging, the atoms are exposed to a standing wave of light that is near-resonant with the $F=2$ to $F'=3$ transition on the D1 line. Given the small detuning, we approximate the system as a two-level atom exposed to a single mode light field, described within the electric dipole and rotating-wave approximations; i.e., the Jaynes-Cummings model \cite{ScullyZubairy}.
To account for the light shifts induced by the dipole trap, we use the position dependent probe detuning  \mbox{$\deltaR$} in the calculation of the dressed states.
Our standard imaging conditions use \mbox{$\delta_\textrm{c}\approx2\pi\times15$~MHz} and an  intensity higher than the on-resonance saturation intensity.
The dressed state energies of the combined light-atom system are \cite{Dalibard1985}
\begin{align}\label{eqn:dressedE}
&E_{1n}(\mathbf{r})=(n+1)\hbar\omega+U(\mathbf{r})-\frac{\hbar\deltaR}{2}+\frac{\hbar\OmegaR}{2},\nonumber\\
&E_{2n}(\mathbf{r})=(n+1)\hbar\omega+U(\mathbf{r})-\frac{\hbar\deltaR}{2}-\frac{\hbar\OmegaR}{2},
\end{align}
where $n$ is the number of photons in the light field.
The generalized Rabi frequency is given by \mbox{$\OmegaR=\left[\deltaR^2+\omega_1^2(\mathbf{r})\right]^{1/2}$},
with the on-resonance Rabi frequency $\omegaR$ in the semi-classical picture defined by \mbox{$\omegaR \rme^{i\phiR}=M_{ij}\cdot E(\mathbf{r})/\hbar$}, where $\phi(\mathbf{r})$ is the phase of the light-atom coupling.
\begin{figure}[t]
\includegraphics[width=8.6cm]{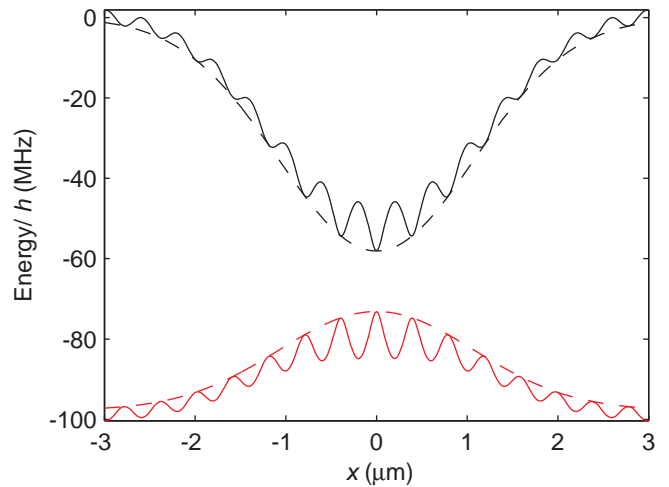}
\caption{(Color online) Calculated dressed state energy for an atom in the microtrap dressed by a standing wave of near-resonant light. The probe beam has power 30~$\mu$W and $\deltac=2\pi\times15$~MHz. The ground state $F=2,m_F=0$ is shifted down by the dipole trap (black dashed line), and modulated by a standing wave of   blue-detuned light on the D1 line (black solid line). The excited state $F'=3,m'_F=0$ is shifted up by the dipole trap (red dashed line), and modulated by the standing wave (red solid line).}
\label{fig:tot_lightshifts}
\end{figure}

Figure~\ref{fig:tot_lightshifts} shows the dressed state energies 
for a $^{85}$Rb atom held in the dipole trap, exposed to the standing wave probe beam with power $30~\mu$W and  \mbox{$\deltac=2\pi\times15$~MHz}.
For simplicity, we treat the probe beam as propagating orthogonal to the dipole trap beam. With the choice of a small blue detuning for an atom at the center of the trap, the effective detuning of the probe beam increases away from the center. Thus, at the center of the trap, the effective probe beam detuning is minimal and the light shift induced by the probe beam is maximal. Away from the center, the effective blue detuning increases and the probe beam light shifts decrease in magnitude. 

\section{Single atom detection}\label{sect:Singleatomdetection}

Our main goal in single atom fluorescence detection is to detect as much light  scattered by the atom as possible while leaving it trapped. 
As such, the two quantities we use to characterize the detection performance are the integrated fluorescence -- the number of ADUs we detect on the camera -- and the retention probability -- the probability that a single atom remains in the microtrap following a fluorescence image.
The retention probability is a measure of the heating caused by the detection process. The higher the single atom scattering rate, the higher the integrated fluorescence signal and the greater the accuracy with which we can distinguish between the presence and absence of an atom  in the microtrap. The integrated fluorescence and the retention probability are coupled in the sense that a low retention probability leads to the loss of an atom from the microtrap during the detection process, and thereby a lower signal than if the atom had remained trapped throughout the exposure.

\begin{figure}[t]
\includegraphics[width=8.6cm]{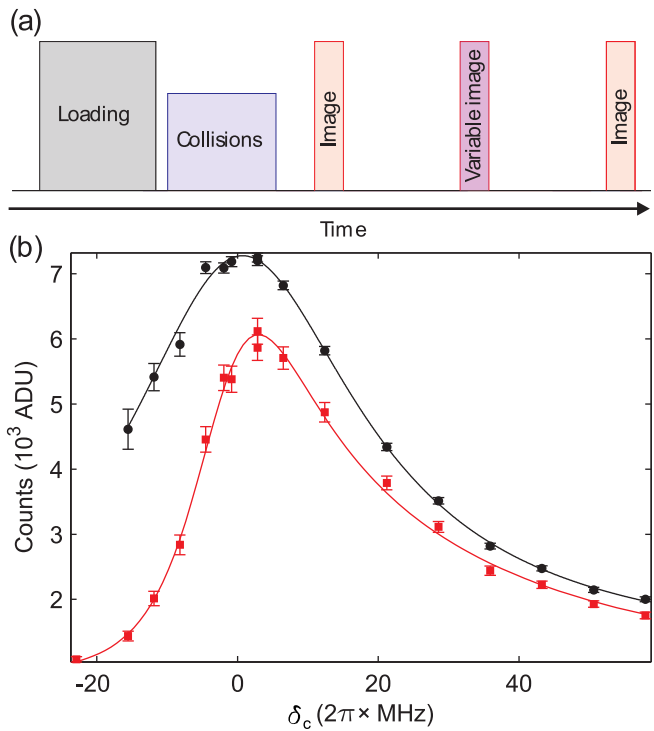}
\caption{(Color online) (a) Experimental sequence to investigate the fluorescence detection of a single atom. (b) Fluorescence spectrum observed when the probe beam detuning is varied. 
Integrated fluorescence conditioned on the detection of a single atom in the first and third image (black circles),   data conditioned only on there being an atom in the first image (red squares). Solid lines are fits of asymmetric Lorentzian functions to the data.}
\label{fig:blue_spectrum_comp}
\end{figure}
Figure~\ref{fig:blue_spectrum_comp}(a) is a schematic of the experimental sequence  we use to investigate fluorescence detection of a single atom held in the microtrap. The sequence begins by loading $\sim100$ atoms into the microtrap from the MOT, whereupon we drive inelastic light-assisted collisions to reduce the number of atoms in the trap to one or zero~\cite{Grunzweig2010,AliciaSingle}.
The first fluorescence image confirms the presence of an atom. For this image,
the probe beam is applied as a standing wave with $30~\mu$W power and detuning  $\delta_\textrm{c}=2\pi\times15$~MHz, and the D2 beams each has power 1.4~mW and detuning $\delta_\textrm{c}^{33'}=2\pi\times2.8$~MHz. 
By this imaging procedure,  an atom is retained in the microtrap with 99\% efficiency. In the second `variable' image, an imaging parameter is varied, and the integrated fluorescence for this setting is recorded.
The third image, identical in parameters to the first, tests whether the atom is still trapped after the second image. The outcome of the third image, conditioned on there being an atom present in the first image, gives the retention probability. All images have 10~ms exposure time. 
Unless otherwise stated, each data point shows the mean of 200 (unconditioned) experimental runs and error bars show the standard error of the mean; to avoid obtaining unphysical retention probabilities exceeding 100\%, we do not correct for the small loss induced by the first and third images when calculating the  probability of retaining an atom. With a view to presenting qualitative arguments to describe single atom detection, we simplify the physical picture by omitting the D1 repump light on the $F=3$ to $F'=2$ transition.  The addition of the D1 repump light does not change the performance of the fluorescence detection other than to increase the observed signal, a feature we make use of when imaging samples containing several atoms (see section~\ref{sect:Multipleatomdetection}). 

To investigate the characteristics of the \bs mechanism as a function of detuning, and to confirm the light shift calculations of section~\ref{sect:LightShifts}, we performed an in-trap fluorescence spectrum of a single atom. To do this, we varied the detuning of the probe beam in the `variable image'  in the experimental sequence (see Fig.~\ref{fig:blue_spectrum_comp}(a)), while holding the power at $30~\mu$W. The MOT cooling beams were set to $\deltacmot=-2\pi\times5~$MHz and 1.4~mW per beam.
Figure~\ref{fig:blue_spectrum_comp}(b) shows the  observed single-atom spectrum.
The   red squares show the mean integrated fluorescence for the experimental runs where there was an atom present in the first image. The black circles show the mean integrated fluorescence conditioned on there being an atom present both in the first and third images. The two data sets have been fitted with an asymmetric Lorentzian model \cite{Stancik200866}. The unconditioned data is asymmetric because the standing  wave probe beam has a cooling effect on the blue side of the spectrum and a heating effect on the red side. This sign dependence of the detuning also shifts the peak of the unconditioned data approximately 3~MHz to the blue. 
In contrast, the peak of the conditioned data agrees  well with $\deltac=0$, confirming the validity of the light shift calculation.  This illustrates a great advantage of non-destructive imaging at the single atom level: It allows one to identify artifacts   from features.  In this case, we avoid the corruption of the spectroscopic signal  due to atom loss. 


The Sisyphus  mechanism realized by the standing wave probe beam cools an atom for blue detunings and heats for red detunings \cite{Dalibard1985}, so it is somewhat surprising that for modest red detunings of the probe beam, \mbox{$-2\pi\times15\lesssim\deltac <0$~MHz}, an atom remains trapped in some  of the experimental runs. There are two reasons for this, both relating to the position of the atom in the microtrap and the effect this has on the frequency of the applied light fields for probing and optical pumping. First, the effective probe beam detuning is pushed to the blue as an atom travels into the wings of the microtrap. As such, although the standing wave probe beam heats an atom at the center of the trap for \mbox{$\deltac\approx -2\pi\times15$~MHz}, in the wings the effective blue detuning is restored and the atom experiences the \bs cooling mechanism, thereby remaining trapped.
Second, the MOT cooling beams laser cool a trapped atom when it moves to the wings of the microtrap: Although tuned so as to optically pump an atom at the center of the trap back to the $F=2$ ground state, i.e., $\delta_\textrm{c}^{33'}\approx 0$, away from the center of the microtrap, an atom experiences the MOT cooling beam frequency closer to the $F=3$ to $F'=4$ cycling transition, thereby realizing a standard red-detuned optical molasses. As such, an atom can be heated at the center of the trap by the probe beam, but be cooled by the MOT beams in the wings of the microtrap so that it remains trapped.

\begin{figure}[t]
\includegraphics[width=8.6cm]{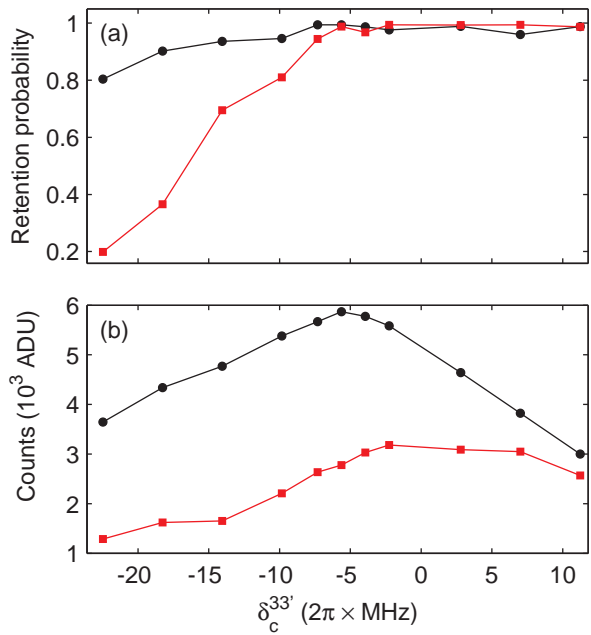}
\caption{(Color online)
Imaging performance as a function of MOT cooling light detuning
 from $F=3$ to $F'=3$,  $\deltacmot$, at constant MOT cooling light power. The experiment was performed using standing wave (black circles) and traveling  wave (red squares) probe beams.  (a) Probability of retaining an atom in the microtrap.  (b) Integrated fluorescence.}
\label{fig:retrovtravel_D2}
\end{figure}

This shows that it is important to characterize the role the MOT cooling light plays in the detection process, even though its nominal role is to optically pump a trapped atom to the electronic ground state addressed by the probe beam. To do this, we varied the detuning of the MOT cooling light while keeping the D1 imaging parameters constant. The experiment was performed for standing and traveling wave probe beams in the second image in order to test the cooling effect of the MOT cooling light in the presence and absence of the Sisyphus cooling mechanism.  As noted above, the blue Sisyphus cooling mechanism relies on the spatial modulation of the probe beam intensity: In a traveling wave configuration, the probe beam has  a flat intensity distribution over the extent of the microtrap and therefore provides no cooling. Figure~\ref{fig:retrovtravel_D2}(a) shows the probability of detecting  an atom in the third image, conditioned on it being detected in the first. For both standing and traveling wave detection and $\deltacmot\geq-2\pi\times5~$MHz, the retention probability is in excess of 97\%. However, below this detuning, both standing and traveling wave probe beams show an approximately linear decrease in retention probability, but the decrease is much sharper for the traveling wave case. Thus, for $\deltacmot\geq-2\pi\times5~$MHz, the D2 beams perform enough laser cooling to keep an atom trapped while imaging for this choice of probe beam parameters, whereas below this detuning, the blue Sisyphus cooling mechanism of the probe beam is necessary to keeping the atom trapped. Figure~\ref{fig:retrovtravel_D2}(b) shows the integrated fluorescence counts of the second image. In the standing wave case, the counts are maximized for  $\deltacmot=-2\pi\times5~$MHz.

\begin{figure}[t]
\includegraphics[width=8.6cm]{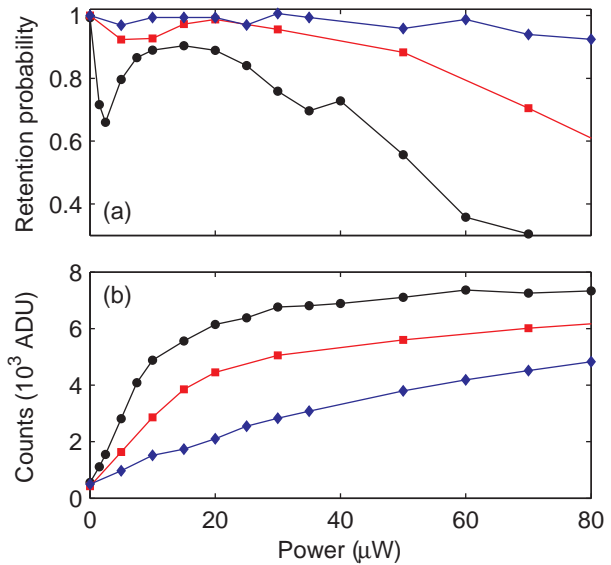}
\caption{(Color online) Single atom detection as a function of probe beam power for three detunings $\delta_\textrm{c}=2\pi\times$ 5 (black circles), 15 (red squares) and 35 (blue diamonds)~MHz. (a) Probability of retaining an atom in the microtrap.  (b) Integrated fluorescence.
}
\label{fig:imaging_D2_1}
\end{figure}
To investigate this imaging process further, we scanned the probe beam power for several values of detuning.
In order to focus on the cooling effect of the \bs mechanism, the D2 light was set to $\deltacmot=-2\pi\times10$~MHz 
with 1.4~mW per beam. This value of the detuning was a compromise between accentuating the cooling effect of the standing wave probe beam, and ensuring efficient optical pumping and hence a good signal.
Figure~\ref{fig:imaging_D2_1} shows the retention probability and integrated fluorescence for three values of the probe beam detuning: \mbox{$\delta_\textrm{c}= 2\pi\times$~5, 15 and 35~MHz.}

 The retention probability shows an interesting structure for the low detuning \mbox{$\delta_\textrm{c}= 2\pi\times~5$~MHz}, which is approximately equal to the natural linewidth \mbox{$\Gamma_\textrm{D1}= 2\pi\times~5.75$~MHz} of the transition. For zero power, the retention probability is of course approximately one, limited only by the performance of the first and third images. However, as the power of the probe beam is increased very slightly, the retention probability drops sharply and reaches a local minimum at a power of  $2.5~\mu$W, corresponding approximately to  saturation intensity $I_\textrm{sat}$ for the transition.  As the power is further increased, the retention probability increases to a peak value and eventually begins to decrease. The initial drop at low power can be understood as Doppler heating predominating over the Sisyphus mechanism. To give a  qualitative picture of this behaviour, we argue using the simple 
 treatment of Doppler heating/cooling, where the total force on an atom due to a counter-propagating pair of laser beams is given by the sum of the contribution from each beam~\cite{Metcalf1999}. Using this description, the Doppler heating rate attains its maximum value for  $\delta_\textrm{c}=\Gamma_\textrm{D1}$ and $I=I_\textrm{sat}$.
Above $2.5~\mu$W, the Doppler heating rate decreases and the cooling effect of the \bs mechanism increases as the magnitudes of the dressed state energies increase at the anti-nodes of the standing wave. The minimum temperature achievable with the \bs mechanism is proportional to the modulation amplitude of the dressed states $E_\textrm{mod}$ in the standing wave \cite{Wineland1992}; as the power is further  increased, the temperature of the sample increases  and eventually becomes commensurate with the trap depth, so that the D2 cooling beams can no longer provide enough cooling to keep the atom trapped. 
If one assumes a Maxwell-Boltzmann distribution for the energy of an atom with temperature $T\sim E_\textrm{mod}/K_\textrm{B}$ in the microtrap,  integrating the distribution from zero to $U_0$ shows the same trend as this experimental retention probability.

At higher blue detunings, the dip in retention probability at low power is significantly reduced: \mbox{$\delta_\textrm{c}= 2\pi\times~	15$~MHz} shows a small decrease at $10~\mu$W, and at \mbox{$\delta_\textrm{c}= 2\pi\times~	35$~MHz} the retention probability is essentially flat for low power. 
This can be understood from the fact that for detunings larger than the natural linewidth, the peak Doppler heating rate shifts to higher intensities, where the modulation of the dressed state energies (Eq.~\eqref{eqn:dressedE})   is already significant, and thus the cooling effect of the \bs mechanism predominates over Doppler heating.

\section{Multi-atom detection}\label{sect:Multipleatomdetection}

In Fig.~\ref{fig:blue_spectrum_comp}, we saw that trap loss can distort the fluorescence signal of a single atom. However, due to the fact that one atom can be detected with near-unit efficiency, this loss effect could be eliminated by adding additional pictures to the experimental sequence and appropriate post-processing. In the case of fluorescence detection of multiple atoms in a microscopic volume, the process is complicated by rapid two-body loss due to light-assisted collisions between atoms. 
In previous work~\cite{McGovern2011}, we showed that fluorescence imaging using  a blue-detuned standing wave of light could be used to count  three atoms held in a microtrap based on the unambiguous signature of steps in the integrated fluorescence. However, this signature is lost when counting larger atomic samples because  light-assisted collisions reduce the atom number during a given exposure time, thereby blurring the steps in fluorescence.
Thus, to accurately count multiple atoms in a microscopic volume, we must devise additional techniques to mitigate the effects of loss during the imaging sequence.

In our system, inelastic light-assisted collisions during fluorescence imaging are the dominant loss mechanism. 
In general, binary collisions between cold alkali atoms in their electronic ground states  occur through an induced dipole-dipole interaction, a second order effect with interaction energy given by  $C_6/R^{6}$, where $R$ is the inter-atom distance and $C_6$ is  the van der Waals interaction coefficient \cite{Weiner2003}. 
In the presence of near-resonant light,  the electronic excited state population is non-zero and atoms can interact via a resonant dipole interaction $C_3/R^{3}$. One can estimate the inter-atom distance  at which the light is resonant with the two-atom system, 
  the Condon point $R_\textrm{C}$,  by $\hbar\delta=C_3/R_\textrm{C}^{3}$. A typical $C_3$ value is 20~eV\AA$^3$, so for $\delta=2\pi\times15~$MHz, $R_\textrm{C}\sim 500$~\AA. For 100 atoms in the microtrap at $160~\mu$K, the peak density is $n_0\sim 5\times10^{13}$~cm$^{-3}$, and the mean inter-atom distance is $R\sim n_0^{-1/3}\sim$250~\AA. Thus, in this system the mean inter-atom distance is of the same order as the interaction length, and light-assisted collisions predominate.

Our approach to counting atoms in  mesoscopic samples is to model the effect of two- and one-body  loss on the time-resolved integrated fluorescence  and from this determine the atom  number as a function of imaging time. A similar  approach using standard red-detuned cooling light was recently demonstrated for up to $\sim 20$ atoms in a larger volume dipole trap~\cite{PhysRevLett.112.043602}. We model the atom loss during imaging with a master equation, in which the time evolution of the probability $p_N$ to obtain $N$ atoms in the microtrap is given by  \cite{PhysRevA.85.035403,vanKampen}
\begin{equation}\label{eqn:pN}
\frac{dp_N}{dt} = \beta(\mathbb{E}^2-\ind)\left[\frac{N(N-1)}{2}p_N\right]+\gamma (\mathbb{E}-\ind)[Np_N],
\end{equation}
where $\beta$ and $\gamma$ are the two-body and one-body decay rates, respectively, and $\mathbb{E}$ is the step operator. This operator is defined by its effect on a function $f(N)$ by \mbox{$\mathbb{E}[f(N)]=f(N+1)$} and  \mbox{$\mathbb{E}^{-1}[f(N)]=f(N-1)$}. The first term in Eq.~\eqref{eqn:pN} describes a two-body loss process, where both atoms are lost from the trap; we call this ``process 2''. One can also model a two-body collision where only one atom is lost from the trap (``process 1'' ) by replacing $\mathbb{E}^2$ in this term by $\mathbb{E}$; in our experiments, such a process arises from the combined effect of the  imaging beam and the MOT cooling beams used to optically pump the atoms. 
 By probing an atomic sample containing just two atoms, we find that process 1 and process 2 occur with approximately equal probability.  Equation~\eqref{eqn:pN} comprises  a system of equations that may be  solved numerically for samples containing several hundred atoms. 

The time evolution of the mean number of atoms, \mbox{$\bar{N}=\sum^\infty_{N=0} Np_N$}, may be found from Eq.~\eqref{eqn:pN}, yielding
\begin{equation}\label{eqn:dNdt}
\frac{d\Nbar}{dt} = -\beta \Nbar (\Nbar-1)-\gamma \Nbar-\beta\Delta N^2,
\end{equation}
where  $\Delta N^2$ is the variance in the atom number. Equation \eqref{eqn:dNdt} also describes process 1, but in this case the first term on the right hand side reads $-\frac{\beta}{2} \Nbar (\Nbar-1)$, given that only one of the two atoms is lost from the trap. Accordingly, to achieve the same mean decay rate when comparing the two processes, we use decay rate $\beta$ for process 2, and $2\beta$ for process 1.  

\begin{figure}
\includegraphics[width=8.6cm]{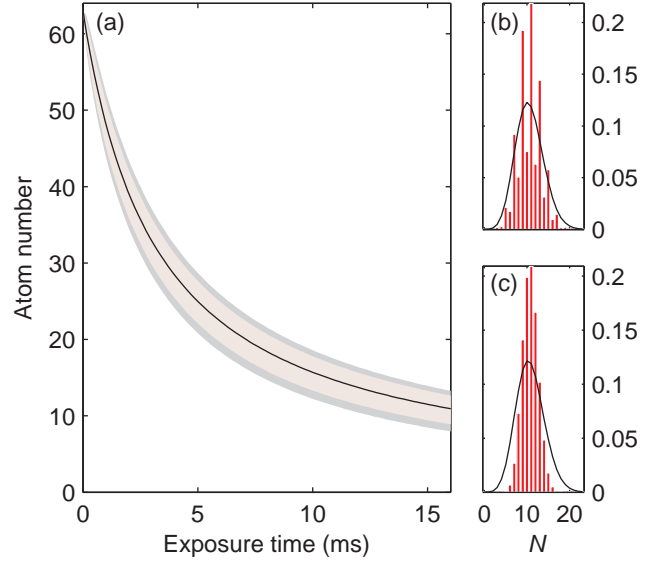} 
\caption{(Color online) Atom number distributions evaluated   for processes  1 and 2 from Eq.~\eqref{eqn:pN} using $N_0=63$, $\beta=4.9$~s$^{-1}$ and  $\gamma=0.97$~s$^{-1}$. To ensure the same mean atom number for the two processes, we use  $\beta$ for process 2 and $2\beta$ for process 1 (see Eq.~\eqref{eqn:dNdt} and associated text). (a) Mean atom number (black line) calculated from Eq.~\eqref{eqn:dNdt} assuming $\beta\Delta N^2=0$. The shaded regions indicate the $2\sigma$ width of the atom number distributions for process 2 (gray) and 1 (red). (b) Atom number distribution at 16~ms for process 2. The solid line shows the poissonian distribution for the mean atom number \mbox{$\bar{N}=10.6$}. (c) Same as (b) but for process 1.
}
\label{fig:Master}
\end{figure}

Figure~\ref{fig:Master} shows the time evolution of the atom number distribution calculated using Eq.~\eqref{eqn:pN} for representative  parameters used in our experiments. Beginning from 63 atoms in the trap, the imaging leads to a decrease in the mean number of atoms and an increase in the atom number variance. 
Figure~\ref{fig:Master}(a) shows $\bar{N}$ as a function of exposure time and  the shaded regions denote the $2\sigma$ width of the atom number distribution for processes 1 and 2.  The solid line shows the mean atom number calculated from Eq.~\eqref{eqn:dNdt} and assuming $\beta\Delta N^2=0$; it is clear that this solution approximates the means of the two distributions very well, showing that this correction to the mean atom number  is negligible for our experimental parameters. 
The atom number variance initially grows from its (idealized) value of zero at $t=0$ to reach its ``limiting'' value at $t\approx10$~ms. This value is determined by the competition between the one and two-body loss mechanisms: Acting alone, the former leads to a poissonian number fluctuations, i.e.,  $\Delta N^2=\bar{N}$, whereas process 2 induces an atom number variance of $0.667 \bar{N}$ for $\bar{N}\gtrsim10$~\cite{PhysRevLett.104.120402}, reducing to $\Delta N^2=0.5\bar{N}$ for the special case of $\bar{N}=0.5$~\cite{Schlosser2001}. Process 1, on the other hand, leads to zero atom number variance  in the limit that $\bar{N}\rightarrow 1$, because the reduction in atom number stops when there is only one atom left in the trap. The differences in the atom number variance produced by  these two-body processes are evident in Figs.~\ref{fig:Master}(b) and (c), which show the atom number distributions at $t=16$~ms for process 2 and 1,    respectively. Process 1 induces a smooth, unimodal distribution, whereas process 2 yields a jagged distribution with a unimodal envelope. For this calculation, odd atom numbers are more probable because the initial atom number was chosen to be odd, and process 2 leads to pairwise loss.
In both cases, however, the width of the distribution is smaller than a poisson distribution with the same mean.

\subsection{Mean values}\label{subsect:meanvalues}
This discussion shows that  to  determine  the mean number of atoms in the trap, we may solve Eq.~\eqref{eqn:dNdt} under the condition that  the atom number variance term is neglected. We will return to the fluctuations about the mean atom number in section \ref{subsect:fluctuations}, where we quantify the precision of the  imaging method. As noted above, the choice of process 1 or 2 does not affect the description of the mean atom number other than the fitted  value of $\beta$.
To simplify the comparison of our method with standard fluorescence detection methods, we model the mean value data with process 2.   
The solution of Eq.~\eqref{eqn:dNdt} within the approximation that 
$\beta\Delta N^2=0$ is  given by
\begin{equation}\label{eqn:Nsoln}
\bar{N}(t) = \frac{\rme^{At}}{\frac{\beta}{A}(\rme^{At}-1)+\frac{1}{N_0}},
\end{equation}
where $A=\beta-\gamma$ and $N_0$ is the initial atom number.
In an experiment, we detect the amount of light scattered by the atomic sample
\begin{equation}\label{eqn:Ftotdef}
F_\textrm{int}(t) = \int_{0}^{t} F_1 \bar{N}(t') dt',
\end{equation}
where we have assumed that the integrated fluorescence is given by the product of the  single atom fluorescence rate $F_1$, which is assumed to be constant in time, and the number of trapped atoms. Performing the integration, we obtain
\begin{equation}\label{eqn:Ftotsoln}
F_\textrm{int}(t) =  \frac{F_1}{\beta} \ln\left[\frac{\beta N_0(\rme^{At} -1)}{A}+1 \right].
\end{equation}
Thus, to determine the initial atom number of a trapped sample, we take in-trap fluorescence images  for several values of exposure time and use Eq.~\eqref{eqn:Ftotsoln} to model how the integrated fluorescence changes in time.   
\begin{figure}[t]
\includegraphics[width=8.6cm]{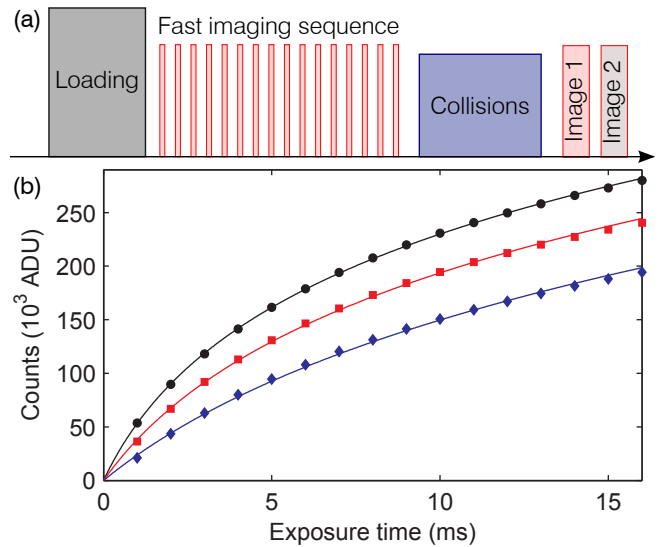}
\caption{(Color online) (a) Experimental timeline for multi-atom detection.
(b)  Integrated fluorescence as a function of  exposure time  for three values of the initial atom number $N_0$. Each data set was  fitted  with Eq.~\eqref{eqn:Ftotsoln} using $F_1$ and $\gamma$ as  fixed input parameters (see text). The fitted parameters were: (black circles) $N_0$=85.9(5), $\beta=6.41(4)$s$^{-1}$; (red squares) $N_0$=58.4(4), $\beta=6.00(5)$s$^{-1}$; (blue diamonds) $N_0$=31.3(3), $\beta=5.70(9)$s$^{-1}$. The number in parenthesis gives the uncertainty from the fitting routine in the least significant figure.}
\label{fig:Natomsingleatomcomp}
\end{figure}

Figure~\ref{fig:Natomsingleatomcomp}(a) is a schematic of the experimental sequence used to determine the number of atoms held in the microtrap. We achieve a given initial atom number $N_0$  by varying parameters in the MOT and CMOT phases; through this loading procedure and evaporation, the atomic clouds thermalize to a temperature $T\sim160~\mu$K
~\footnote{For instance, to load a small number of atoms into the microtrap, we shorten the MOT loading time to $\sim500$~ms,  and reduce the magnetic field gradient and the duration of the CMOT phase. This procedure leads to a shot-to-shot variation of about 10\% in the loaded atom number once the experiment is thermally stable. Alternatively, one can leave the experimental parameters for loading constant  and then drive inelastic light-assisted collisions to reduce the atom number. However, by this method, the temperature of the atomic sample is increased.}. 
The time dependence of \Fint  is probed by exposing the trapped sample to several short pulses of imaging light and recording  the fluorescence on the EMCCD camera. For this fast imaging sequence, the camera is operated in  kinetics mode and the camera chip masked such that we can take $\sim 20$ frames for our chosen region of interest. Each imaging pulse has 1~ms duration, with a 22~ms delay between frames. \Fint is generated by taking the cumulative sum of the integrated signal for each frame \footnote{Note that by working with the cumulative signal, the method is also applicable using a camera operated in a standard mode, i.e.,  a camera without the ability to take several images in quick succession.}.
In order to measure $F_1$, at the end of the fast imaging sequence we drive light-assisted collisions with the probe beam set to reduce the trapped atom number to zero or one \cite{Grunzweig2010,AliciaSingle}, whereupon we take a  10~ms fluorescence image (``Image 1'' in Fig.~\ref{fig:Natomsingleatomcomp}(a)) using the same parameters as for the fast imaging sequence.  Finally, a standard image  (``Image 2'') of duration 10~ms 
 determines with near unit efficiency whether there was zero or one atom held in the microtrap in ``Image 1''. Using the final two images,  and averaging over several experimental runs, we obtain the  fluorescence rate for one or zero atoms, allowing us to calculate the background subtracted single atom fluorescence rate $F_1$ and the background subtracted integrated fluorescence  $F_\textrm{int}$. The use of collision beam parameters to produce 50\% zeros and ones minimizes the statistical error in the measured background subtracted single atom  fluorescence rate, meaning that one could also employ the standard approach of using \lac driven by red-detuned molasses for this step. 

Figure~\ref{fig:Natomsingleatomcomp}(b) shows the integrated fluorescence as a function of  exposure time  for three values of the initial atom number $N_0$. The experiment was performed for a dipole trap power of 30~mW, yielding a trap depth of 
 \mbox{$U_0  = h\times87.1$~MHz~=~$\Kb\times 4.18$~mK}, with 
standing wave probe beam  parameters $\deltac=2\pi\times15$~MHz and power $40~\mu$W for both the $F=2\rightarrow F'=3$ and  \mbox{$F=3\rightarrow F'=2$} D1 transitions,  and MOT cooling beam parameters \mbox{$\deltacmot=-2\pi\times6$~MHz} at 0.6~mW per beam.
 The data has been fitted with Eq.~\eqref{eqn:Ftotsoln}, using $F_1$ and $\gamma$ as fixed parameters. The single atom decay rate was measured to be $\gamma= 0.97$~s$^{-1}$ in a separate experiment where a single atom was prepared in the microtrap and exposed to the same imaging conditions as we use for $N$-atom samples. The initial slope of the integrated fluorescence for each data set is given by $N_0F_1$. 
  For longer exposure times, the   fluorescence rate decreases as a result of the loss of atoms from the trap during imaging. It is evident that Eq.~\eqref{eqn:Ftotsoln} models the experimental data very well, yielding measured  initial atom numbers of 31, 58 and 86 for the three data sets. We have verified the method for initial atom numbers up to 130, which is approximately the highest number of atoms  we obtain in the microtrap using the described loading procedure.
 
The measured values of two-body loss rate are significantly smaller than those reported for comparable experiments that use red-detuned imaging light. Our low value of $\beta$ arises from the fact that the probe beam is blue-detuned with respect to the primary transition used for imaging: The blue detuning excites the colliding atoms onto a repulsive molecular potential curve \cite{Bali1994}, which limits trap loss due to inelastic light-assisted collisions by ensuring that the maximum energy the atom pair gains is $\hbar\delta$; for our experimental parameters, this gained energy is significantly smaller than the trap depth. In order to compare the measured values of $\beta$ with those from other experimental configurations, we calculate the normalized two-body loss rate $\beta_\textrm{norm}=\beta 2 \sqrt{2}V$, where \mbox{$V=(2\pi\Kb T/(m\bar{\omega}^2))^{3/2}$} is the volume occupied by the sample  at temperature $T$, $\bar{\omega}$ is the geometric mean of the trapping frequencies, and $m$ is the atomic mass \cite{PhysRevA.85.062708}. From Fig.~\ref{fig:Natomsingleatomcomp}, we measure a typical  value $\beta=6$~s$^{-1}$, which in normalized form reads $\beta_\textrm{norm}=1.4\times 10^{-11}$~cm$^3$s$^{-1}$. 
For comparable experiments using red-detuned fluorescence imaging light, the normalized two-body decay rate is typically two orders of magnitude larger~\footnote{See Ref.~\cite{PhysRevA.85.062708} for a brief survey of measured values of $\beta$ found in the literature.}. In particular, we find  $\beta_\textrm{norm}=1.4\times 10^{-9}$~cm$^3$s$^{-1}$ for the experiment described in Ref.~~\cite{PhysRevLett.112.043602}.   It is this feature of our imaging technique that permits the determination of atom number in samples of very high density.

In our treatment, we assume that the single atom fluorescence rate $F_1$ does not depend on the atom number. This assumption may lead to systematic errors for very dense samples. In particular, light scattering may be suppressed due to dipole-dipole interactions \cite{PhysRevLett.113.133602} or if the temperature of the atoms is number dependent. The latter would lead to a number dependent broadening of the transition due to the trap light shifts. An atom number dependent temperature could arise due to heating from inelastic collisions that release a low energy such that the colliding atoms are not lost. Inclusion of 3-body processes in the model (Eqs. \eqref{eqn:dNdt} to \eqref{eqn:Ftotsoln})
 did not improve the fit in Fig.~\ref{fig:Natomsingleatomcomp}(b), but for higher densities they may also play a role. Reference \cite{PhysRevLett.112.043602} demonstrated that the atom number in small samples can be determined from the Rabi frequency of collective Rydberg excitations, which may provide a method for detecting potential systematic errors in our parameter regime. 
 However, in the following section we present a noise model for the imaging process which builds on the fitted values of $N_0$ and $\beta$. This model contains no additional free parameters and shows very good agreement with the experimental data, thereby  supporting the accuracy of the method.

\subsection{Fluctuations}\label{subsect:fluctuations}
In this section, we develop a noise model to characterize the imaging method described above, thereby extending the method from the mean atom number to the  atom number distribution within the microtrap. Similar to previous work \cite{PhysRevLett.109.133603, PhysRevLett.111.253001}, we characterize the noise properties of the imaging method by calculating the two-sample variance from the atom number measured by two consecutive imaging pulses: $\sigma^2=\frac{1}{2}\textrm{var}(N_2-N_1)$. Contrary to previous work, however, we calculate the  noise model based on the full solution of the master equation, and use  no free parameters to fit the noise model to the data. 
We also compare the atom number at the end of the first imaging pulse $N_1$ with the atom number at the beginning of the second imaging pulse $N_2^0$. Such an imaging procedure is experimentally relevant, allowing one to determine the atom number in the microtrap, perform an experiment in which the atom number changes, and finally measure the number of atoms remaining in the trap.
In light of the discussion of Eq.~\eqref{eqn:pN}, one may also view the imaging process as a method to prepare an atomic sample with an atom number defined better than the poissonian limit.

The experimental sequence used to characterize the noise properties of the imaging method is identical to that shown Fig.~\ref{fig:Natomsingleatomcomp}(a), except that in the fast imaging sequence we take only two images, each of duration $\tau$. 
The atom number may be determined from a single imaging pulse by combining  Eqs.~\eqref{eqn:Nsoln} and \eqref{eqn:Ftotsoln}. At the end of the $i$th imaging pulse, the atom number  is given by
\begin{equation}\label{eqn:Ni}
N_i=\frac{A}{\beta}\left[\frac{1-e^{-\beta F_\textrm{int}/F_1}}{1-e^{-A\tau}} \right],
\end{equation}
and at the beginning of the pulse by
\begin{equation}\label{eqn:Ni0}
N_i^0=\frac{A}{\beta}\left[\frac{e^{\beta F_\textrm{int}/F_1}-1}{e^{A\tau}-1} \right].
\end{equation}
The data set for this section comprises 100 experimental runs for each value of $\tau$, with the exposure time  sampled between 0.1 and 3~ms. The value of $\beta$ was determined using the method of the previous section (see Fig.~\ref{fig:Natomsingleatomcomp}(b) and associated text), where the integrated fluorescence for each setting of $\tau$ was found by averaging the signal over the 100 experimental runs; the obtained value was  $\beta=4.94$~s$^{-1}$. 
The initial atom number for the data set was gaussian distributed  with  mean value 62.5 and  a standard deviation of 8.5 due to fluctuations in the loading process; this was obtained from averaging  $N_1^0$ obtained from Eq.~\eqref{eqn:Ni0} over the data set.

\begin{figure}[t]
\includegraphics[width=8.6cm]{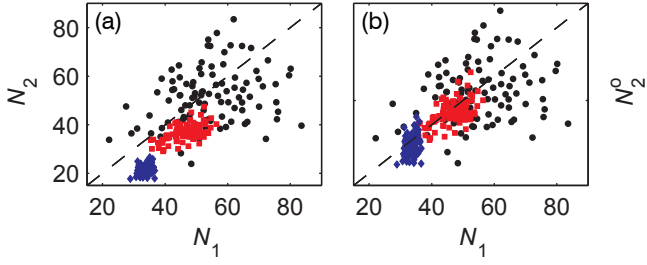}
\caption{(Color online) (a) Measured atom number at the end of pulse 1 and 2 for three imaging pulse durations: $\tau=0.1$~ms (black circles),  $\tau=1$~ms (red squares),  $\tau=3$~ms (blue diamonds). The dashed line shows $N_2=N_1$. (b) Same as (a), but we compare the atom number at the end of the first pulse $N_1$ and the number at the beginning at the second pulse $N_2^0$ inferred from the second pulse.
}
\label{fig:scatter}
\end{figure}

The noise properties of the measured atom number changes as a function of the exposure time.
Figure \ref{fig:scatter}(a) shows the measured atom number $N_2$ v $N_1$ for three values of the pulse duration.  For the shortest pulse duration, the signal shows a large degree of scatter; in this regime, the shot noise of the detected light is the dominant noise source. For longer pulse durations, the scatter decreases markedly but the mean atom loss during the second imaging pulse causes the data to deviate from the line $N_2=N_1$. Figure~\ref{fig:scatter}(b) shows $N_2^0$ v $N_1$. 
Given that $N_1$ and $N_2^0$ are measurements of the same atom number, the data lies on the line $N_2=N_1$. However, the inferred atom number at the start of the second pulse shows a larger scatter than $N_1$.
To understand this behaviour, we model the noise in the measured atom number arising from the shot noise in the detected light and the loss of atoms during the imaging process. 

The light shot noise contribution to the atom number variance may be found through standard error propagation.  
In particular, the variance in the atom number at the end of an imaging pulse due to light shot noise is given by \mbox{$\sigma^2_{N_i}=(dN_i/dF_\textrm{int})^2\sigma^2_{F_\textrm{int}}+(dN_i/dF_1)^2\sigma^2_{F_1}$,} assuming negligible uncertainty in the values of  $\beta$ and $\gamma$. A comparison of the two remaining terms in the error propagation formula shows that the term   proportional to $\sigma^2_{F_1}$  is also negligible.
Thus, $(dN_i/dF_\textrm{int})^2\sigma^2_{F_\textrm{int}}$ gives the variance of the atom number arising from the light shot noise in pulse $i$. 
In principle, $\sigma_{F_\textrm{int}}^2= g^2\eta {N}_\textrm{ph}= gF_\textrm{int}$ (see discussion of the camera in Sec. II).
However, due to the spatial variation in the photon scattering rate for an atom moving in the microtrap and imaging light, and the excess noise induced by the EMCCD camera, $\sigma_{F_\textrm{int}}^2=\alpha F_\textrm{int}$, where $\alpha\approx110$ and is  determined from the histogram of ADU counts for a single atom exposed to the same imaging conditions.
Despite the light shielding of the apparatus, the background signal is also finite so that it  makes a contribution $\sigma_0^2$  to  the variance of the integrated signal: $\sigma_{F_\textrm{int}}^2=\alpha F_\textrm{int} + \sigma_0^2$, where both terms depend on the exposure time. 

 To model atom loss during the imaging pulses, we perform a Monte-Carlo simulation based on the master equation of Eq.~\eqref{eqn:pN}. The atom number distribution at the end of the first imaging pulse is obtained by solving  Eq.~\eqref{eqn:pN} using the initial conditions obtained from the experiment. 
 At time $\tau$, the atom number is sampled and rounded to the nearest integer to give $N_1$; this atom number is then used as the initial condition for the atom number distribution's evolution during the second imaging pulse. The resulting probability distribution is sampled to obtain $N_2$. This procedure is repeated 1000 times for each value of $\tau$, from which we calculate the two-sample variance due to  atom loss, $\sigma_\textrm{Loss}^2$. To evaluate the noise properties of the method when comparing $N_1$ and $N_2^0$, the sampled value of $N_2$ is  inserted into Eq.~\eqref{eqn:Nsoln} and rearranged to yield  $N_2^0$, whereupon we evaluate  the two-sample variance $\frac{1}{2}\textrm{var}(N_2^0-N_1)$.

\begin{figure}[t]
\includegraphics[width=8.6cm]{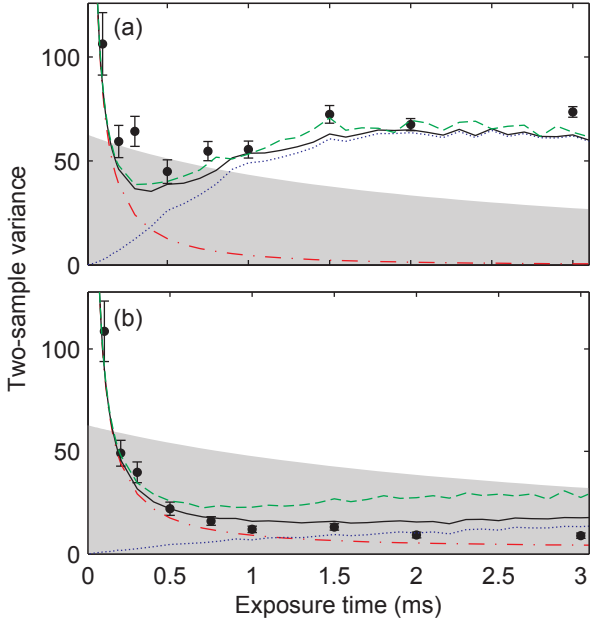}
\caption{(Color online) (a) Two-sample variance of $N_1$ and $N_2$ as a function of exposure time. (Black circles) Experimental data, error bars show $2\sigma$ statistical error, (red dash-dot line) calculated light  noise $\sigma^2_{N_1}+\sigma^2_{N_2}$, (blue dotted line) calculated noise arising from atom loss $\sigma^2_\textrm{Loss}$ based on Eq.~\eqref{eqn:pN} using process 1, (black solid line) calculated total noise \mbox{$\sigma^2=\sigma^2_{N_1}+\sigma^2_{N_2}+\sigma^2_\textrm{Loss}$} using process 1, (green dashed line) calculated total noise model using process 2. The shaded region indicates the interval where the atom number variance lies below the level of poissonian fluctuations.
(b) Same as (a), but for $N_1$ and $N_2^0$.
}
\label{fig:twosample}
\end{figure}

Figure~\ref{fig:twosample}(a) shows the two-sample variance as a  function of exposure time for  $N_1$ and $N_2$. Also shown is the   two-sample variance predicted by the noise model \mbox{$\sigma^2=\sigma^2_{N_1}+\sigma^2_{N_2}+\sigma^2_\textrm{Loss}$}. As in Fig.~\ref{fig:scatter}(a), the scatter of the experimental points initially decreases with the exposure time and reaches its minimum  at $\tau\approx0.4$~ms.  The noise model shows that  the noise contributions from the  light and the  atom loss are approximately equal at this point. For larger $\tau$, the atom loss becomes the dominant noise source and the two-sample variance increases. In a small interval of $\tau$ values around $0.4$~ms, the two-sample variance in detected atom number lies below the level of poissonian fluctuations. 
 For exposure times longer than 3~ms, the two-sample variance again decreases due to two effects arising from the evolution of the atom number distribution: The first stems from the decreased loss rate at late times, meaning that the relative difference in mean atom number $\bar{N}_2-\bar{N}_1$ is smaller than for low $\tau$ (see Fig.~\ref{fig:Master}(a)); the second arises from the reduced variance in $N_1$ and $N_2$ at late times due to   light-assisted collisions (see Figs.~\ref{fig:Master}(b) and (c)). 

The two-sample variance associated with  $N_1$ and $N_2^0$ is shown in Fig.~\ref{fig:twosample}(b). The variance predicted by the noise model is given by \mbox{$\sigma^2=\sigma^2_{N_1}+\sigma^2_{N_2^0}+\sigma^2_\textrm{Loss}$}. In the absence of the atom number variance associated with mean atom loss, the light shot noise is the dominant noise source until $\tau\approx 1$~ms, and remains finite at higher $\tau$, consistent with Fig.~\ref{fig:scatter}(b) where it is apparent that the operation of Eq.~\eqref{eqn:Ni0} to obtain $N_2^0$ leads to increased variance. Nonetheless, for all exposure times above $\tau=0.1$~ms, the imaging technique leads to a two-sample variance well-below the level of poissonian fluctuations. Indeed, for $0.75\leq \tau \leq 3$~ms, the two-sample variance lies a factor of  three to four below the poissonian limit, which is remarkable given the very high density of the atomic sample.

In calculating the two-sample variance for  $N_1$ and $N_2^0$, the use of process 1 in the noise model yields a significantly lower value of noise than the case of process 2, whereas for $N_1$ and $N_2$, the two loss processes give rise to approximately the same noise level. This arises from the fact that the two-sample variance  for $N_1$ and $N_2$ is dominated by the mean atom loss, so that the contribution from the widths of the atom number distributions is small.  In contrast, by comparing $N_1$ and $N_2^0$ we remove the contribution of mean atom loss to the two-sample variance, whereby the choice of loss mechanism becomes the determining factor. As shown in Fig.~\ref{fig:Master}, process 2 leads to a broader number distribution than process 1. Thus, process 2 leads to a larger uncertainty in atom number $N_2$, which is amplified by the inversion of Eq.~\eqref{eqn:Ni0} to obtain $N_2^0$.

The experimental data in Fig.~\ref{fig:twosample}(b) is better matched by the noise model assuming process 1, implying that this process is  the dominant loss mechanism. As noted above, however, an experiment with just two atoms showed that  process 1 and 2 occur with approximately equal probability. A partial explanation for this discrepancy could lie in the heating of the sample due to the high rate of inelastic collisions when probing many atoms in a dense sample: In \cite{PhysRevA.88.051401} we found that an increase in  sample temperature led to a higher probability for process 1 in inelastic collisions that release an energy less than the trap depth. 
Finally, at large exposure times, the experimental two-sample variance lies slightly below that predicted by process 1, indicating 
 that there may be effects that are not  captured by the noise model.

\section{Conclusion and outlook}\label{sect:Conclusion}
We have demonstrated a method of in-trap fluorescence imaging based on a standing wave of light that is blue-detuned from atomic resonance. In general, imaging in red-detuned dipole traps is complicated by the multitude of atomic energy levels and their different light shifts induced by the dipole trap beam(s). Here, given  the use of probe light tuned to the blue of the uppermost hyperfine level,  $F'=3$  on the D1 line of \Rb, the probe has the greatest effect at the center of the trap, meaning that a two-level atom picture is sufficient. 
By combining the standing wave probe beam with a standard set of MOT cooling beams, we achieve a single atom detection probability of approximately 99\% for a wide range of experimental parameters.

We have explored several features of single-atom detection using this method including the role of detuning for the  probe  and MOT cooling beams, and the intensity dependence of the Sisyphus cooling mechanism. The use of single atoms to characterize the method was important to avoid spurious results arising from the single-atom loss caused by the detection process.

The method was extended to the detection of up to $\sim100$ atoms held in the microtrap, at densities  exceeding $10^{13}$~cm$^{-3}$. 
When imaging such high-density samples in a trap of finite depth, inelastic two-body  light-assisted collisions comprise the dominant loss mechanism. We mitigated this loss  
by using a blue-detuned probe beam. The effect of the small blue detuning is two-fold: It reduces the  rate of \ilac via optical shielding, and, when an inelastic collision occurs, it limits the energy gained by the colliding atom pair to the detuning of the imaging light. By choosing the trap depth to be much larger than  the energy gained in an inelastic collision, we achieved very small  two-body loss rates $\beta$ for such a  dense sample. Indeed, the obtained values of the two-body loss rate, normalized to take account of the microscopic volume of the atomic sample, were two orders of magnitude smaller than those obtained using the standard approach of fluorescence detection based on red-detuned molasses light.

We modelled the effects of one- and two-body loss on the sample using a master equation to determine the atom number as a function of imaging time. For two-body loss, we considered two loss ``channels'': Process 1, where one atom is lost from the trap due to a binary light-assisted collision, and process 2, where both atoms are lost. These two processes lead to different atom number distributions in the trap, the signature of which was evident in the noise analysis  of the imaging method. 
The method leads to the preparation and determination of the atom number up to a factor of four below the poissonian limit. When combined with internal state preparation, this method can be used to generate mesoscopic atomic samples useful for quantum information processing. 

\begin{acknowledgments}
This work is supported by the Marsden Fund and UORG. The authors thank T. Gr\"unzweig and M. McGovern for their important contributions at an early stage of the experiment and for comments on the manuscript.
\end{acknowledgments}

%

\end{document}